\documentclass[preprint,preprintnumbers,amsmath,amssymb,showpacs,superscriptaddress]{revtex4}
\usepackage{graphicx}
\usepackage{dcolumn}
\usepackage{bm}

\begin{document}

\preprint{IFIC/16-52}

\title{Can the  $ 750\, GeV$  enhancement be a signal of light magnetic monopoles?}

\author{L.N. Epele,H. Fanchiotti, C.A. Garc\'{\i}a Canal\\
{\it Instituto de F\'{\i}sica La Plata, CCT La Plata, CONICET and Laboratorio de F\'{\i}sica Te\'{o}rica\\
Departamento de F\'{\i}sica,
Facultad de Ciencias Exactas, Universidad Nacional de La Plata\\ CC 67, 1900 La Plata, Argentina }\\
V. A. Mitsou\\
{\it Instituto de F\'{\i}sica Corpuscular\\ 
Consejo Superior de Investigaciones Cient\'{\i}ficas \\ 
Apartado de Correos 22085, E-46071 Paterna (Val\'{e}ncia),
Spain}\\
V. Vento\\
{\it Departamento de F\'{\i}sica Te\'{o}rica and Instituto de F\'{\i}sica Corpuscular
 \\Universidad deValencia-Consejo Superior de Investigaciones Cient\'{\i}ficas
 \\ E-46100 Burjassot (Val\'{e}ncia),
Spain}}
\date{\today}

\begin{abstract}

The announced $\sim 3\,\sigma$ enhancement in the inclusive $\gamma \,\gamma$-spectrum at $\sim 750\, GeV$ made
 by the ATLAS and CMS collaborations at LHC might indicate the existence of a monopole-antimonopole bound state: monopolium. In here we revisit our calculation of 2012 from a more general perspective and see that this resonance, if confirmed,
 might be  a first signal of the existence of magnetic monopoles.

\end{abstract}


\maketitle

\section{Introduction}

Inspired by the old idea of Dirac and Zeldovich \cite{Dirac:1931kp,Dirac:1948um,Zeldovich:1978wj} that monopoles are not seen because they are hidden by their strong magnetic forces forming a bound state called monopolium we proposed some time ago, that due to its bound state structure, monopolium  might be easier to detect than free monopoles  \cite{Epele:2007ic,Epele:2008un}.
   Some years ago \cite{Epele:2012jn}, we stated that the Large Hadron Collider was an ideal machine to discover monopolium and monopole-antimonopole pairs with monopole masses below $1\,TeV$. We proposed the observability of monopoles and monopolium  in the $\gamma \,\gamma$ channel for monopole masses in the range $500\, \sim \,1000\, GeV$ at present running energies and for an integrated luminosity of $5 fb^{-1}$. It is therefore natural to consider the interpretation of the recent $\sim 3\,\sigma$ enhancement in the inclusive $\gamma \,\gamma$- spectrum at $\sim \, 750\, GeV$ reported by the ATLAS \cite{ATLAS,Aaboud:2016tru}
and CMS \cite{CMS,Khachatryan:2016hje} collaborations as a possible monopolium state decaying into its natural diphoton final state. In here we are going to revisit the previously mentioned reference adjusting the monopole mass and the binding energy to reproduce the announced experimental scenario.

Due to the large monopole-photon coupling there is no well defined field theory able to describe the coupling of photons to monopoles. Two schemes have been used thus far: the original Dirac coupling \cite{Dirac:1948um} and the $\beta$-coupling  \cite{Epele:2012jn,Ginzburg:1982fk,Milton:2006cp}. In the former 
the coupling is described directly by $g$ the magnetic charge of the monopole. In the latter the coupling is given by $\beta g$, where $\beta$ is the velocity of the monopoles or monopolium in the two photon center of mass system depending on the process.The Dirac scheme has the largest possible coupling at threshold, while the $\beta$-coupling has the smallest coupling at threshold, namely there it vanishes. In our dual QED theory monopolium can decay not only by $2\gamma$ but also by $4 \gamma$, $6 \gamma$, $\ldots$ sizeably because the coupling is strong. We will parametrize these other decays  by a branching ratio to $2 \gamma$  which we obtain from the LHC data analysis \cite{Harland-Lang:2016qjy,Fichet:2015vvy,Csaki:2016raa}. 

In the next section we  discuss monopolium decay in the two schemes. One immediate consequence of the existence of monopolium is that also monopole-antimonopole can annihilate from their mass threshold onward. This process is studied in Section III. Finally some conclusions of our analysis are drawn in Section IV.

\section{Monopolium decay}

The calculation follows the procedure and approximations developed in ref.\cite{Epele:2012jn} with minor changes. The mechanism proposed to produce monopolium is photon fusion \cite{Kurochkin:2006jr}. We consider all possible processes, namely elastic, semielastic and inelastic. The elementary cross section to be plugged in the formalism of p-p collisions \cite{Drees:1994zx} is given by

\begin{equation}
\sigma (\gamma\, \gamma \rightarrow M \rightarrow \gamma \, \gamma )  = \frac{4\pi}{E^2}  \frac{M ^2
\,\Gamma(E) \Gamma_{\gamma \gamma}}{\left(E^2 - M^2\right)^2 +
M^2\,\Gamma_{Tot}^2}. \label{ppM2}
\end{equation}
Here $M$ is the monopolium mass, $E$ the center of mass energy of the subsystem $\gamma \gamma \rightarrow M \rightarrow \gamma \gamma$ with on shell photons as in the Weizs\"acker-Williams approximation \cite{WW}.  In the Breit-Wigner approximation we take $\Gamma_{Tot} \sim 50$, GeV  and the branching fraction to 2 photons we take of a few percent \cite{Harland-Lang:2016qjy,Fichet:2015vvy,Csaki:2016raa}.

The value of $\Gamma(E)$ subsume the structure of our proposed intermediate bound state and is given by \cite{Epele:2012jn}

\begin{equation}
\Gamma(E)= 2\left(\frac{ \beta_s^2}{\alpha}\right)^2 \left(\frac{m}{M}\right)^3  \left(2- \frac{M}{m}\right)^{3/2} \; E.
\end{equation}
Here, $\alpha$ is the fine structure constant, $m$ is the monopole mass and $\beta_s$ a scheme dependent factor:  $1$ in the Dirac scheme and the monopolium velocity $\beta(M)=\sqrt{1-\frac{M^2}{E^2}} $ in the $\beta$-scheme. The details of the calculation of $\Gamma(E)$ are described in our previous work \cite{Epele:2012jn}, to which we have incorporated the energy dependence \cite{Cooper:1988km} in the required  Weizs\"acker-Williams approximation.
Introducing these elementary expressions in the p-p scattering formalism developed for this problem  one obtains

\begin{equation}
\frac{d\sigma(E)}{dE} = \frac{2 E}{s} \; \sigma (\gamma\, \gamma \rightarrow M \rightarrow \gamma \, \gamma ) \; \mathcal{L}(E),
\label{xsection}
\end{equation}
where $E$ is the center of mass energy of the subsystem and $s$ is the center of  mass energy squared of the p-p system. The last factor following the notation of  ref.\cite{Drees:1994zx} for the inelastic channel is given by
\begin{equation}
\mathcal{L}(E) = \int_{\frac{M^2}{s}}^1 du \int_{\frac{M^2}{s}}^u dt \int_{\frac{M^2}{s}}^t dr \frac{1}{r \,t^2\,u} F(r) F(\frac{t}{r}) f(\frac{u}{t}) f(\frac{v}{u}),
\end{equation}
after having performed the change of variables  $r=x_1$, $t=x_1 x_2$, $u=x_1 x_2 z_2$ and $v= x_1 x_2 z_1 z_2 $ and having used  $s= v E^2$. Analogouss expressions  for the semielastic and elastic channels can be obtained by a similar change of variables. In this way the elementary cross section appears always separated from the convolution integrals.

\begin{figure}
[ptb]
\begin{center}
\includegraphics[
height=2.5in,
width=4in
]%
{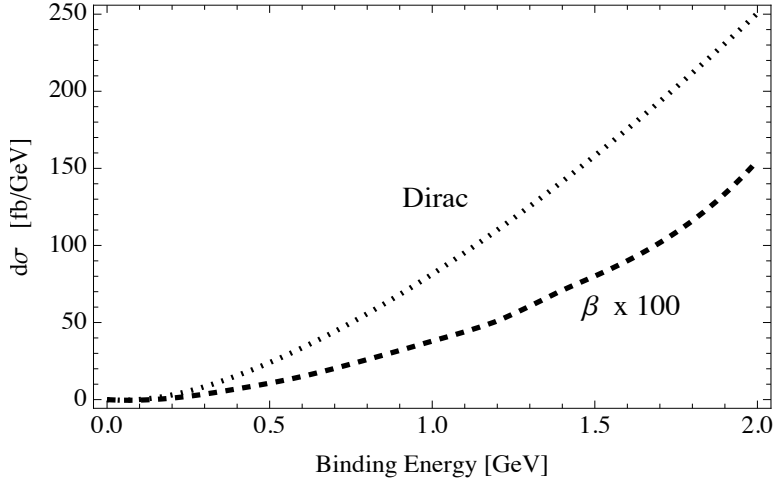}%
\caption{The dasehd line represents the value of the $ 2\, \gamma$ cross section (x100) for the beta coupling, while the dotted line is that corresponding to the  Dirac coupling as a function of monopole mass.}%
\label{DiracvsBeta}%
\end{center}
\end{figure}

We proceed to calculate the cross-section, namely the sum of the three contributions elastic, semielastic and inelastic, as a function of monopole and monopolium masses for the two schemes. We perform the calculation for a total center of mass energy of 8 and 13 TeV and fix our parameters $M$ and $m$ such that their values  lead to a possible resonant  behavior at $750\,$ GeV using  $\Gamma_{Tot}=\,50$ GeV and $\Gamma_{\gamma \gamma} = \, 1$GeV.  The total width and the branching ratio are crucial to determine the precise value of the cross section. A larger width and a smaller branching ratio would diminish the cross section for fixed monopole mass.

We show in Fig.\ref{DiracvsBeta} the cross section for both schemes  as a function of binding energy for a total p-p center of mass energy of $13$ TeV.
Fixing the enhancement at $750\, GeV$ and limiting the cross section to be below $100$ \,fb/GeV leads in the Dirac scheme to strong monopole mass and binding energy restrictions, namely  binding energies can only go up  to $1\, $GeV and monopole masses  have to be below $360\,GeV$. These restrictions imply that the monopole-antimonopole threshold is very close to monopolium and moreover for such a small mass, as will be shown in the next section, even if the monopolium peak is undetectable at $8$ TeV, the monopole-antimonopole annihilation cross section will appear at these energies as a clearly detectable rise of the background. The fact that no such rise has been observed in the first LHC run at these energies implies that the $750$  GeV enhancement cannot be due to monopoles if the conventional Dirac scheme is the one chosen by nature. We assume that that is not the case and proceed to analyze the results in the  $\beta$ scheme from now on.

\begin{figure}
[ptb]
\begin{center}
\includegraphics[
height=2.5in,
width=4.in
]%
{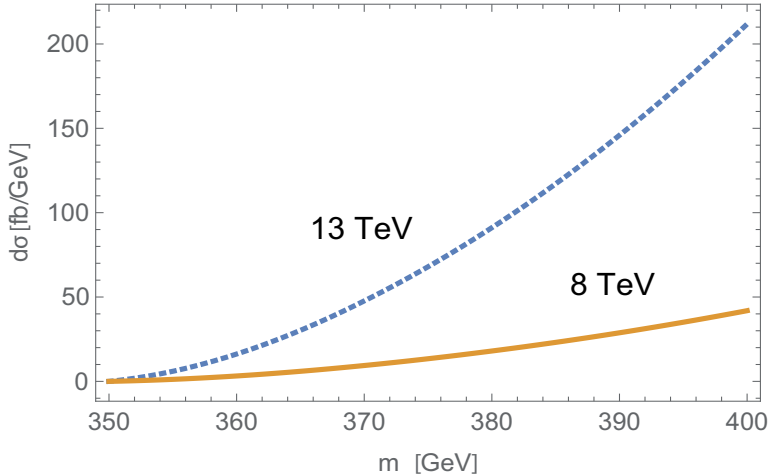}%
\caption{The 2 $\gamma$ cross section  for the $\beta$-coupling scheme as a function of monopole mass for an extended mass range at a  pp center of mass energy of $8$ TeV (solid)  and of $13$ TeV (dashed).}%
\label{13-8}%
\end{center}
\end{figure}

 In Fig \ref{13-8} we plot the cross section as a function of monopole mass for $8$ and $13$ TeV showing that the  $\beta$-coupling is less restricted by the enhancement conditions  and therefore the possible monopole masses can reach higher values ($m <400\, $GeV). In this case the binding energy can be larger reaching values of  up to   $100 \,$GeV at the highest cross sections.  This result is very appealing because it implies that above the suspected enhancement, monopole-antimonopole pairs should be detected in the two photon channel. The $8$ TeV result is about 5 times smaller that the $13$ TeV result in the mass range considered a result which is consistent with the absence of events in the previous LHC run.

\begin{figure}
[ptb]
\begin{center}
\includegraphics[
height=2.5in,
width=4.in
]%
{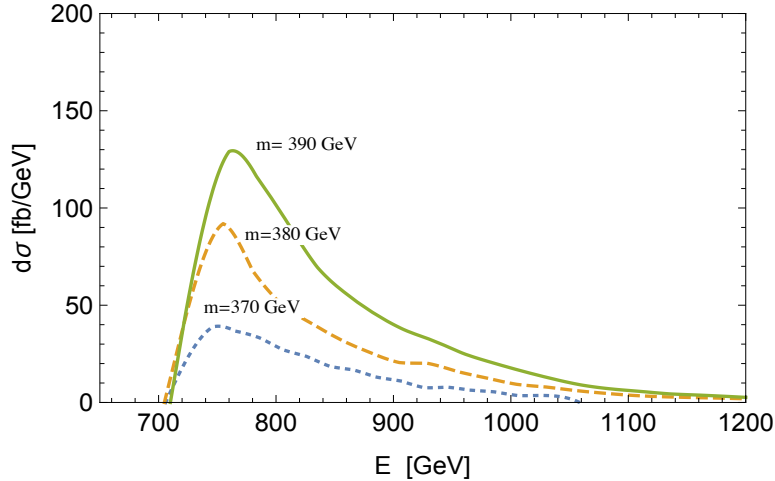}%
\caption{The 2 $\gamma$ cross section  for monopolium decay in the  the $\beta$-coupling scheme for several monopole masses  as a function of the two photon center of mass energy.}%
\label{monopolium}%
\end{center}
\end{figure}

In Fig. \ref{monopolium}  we show the resonance peak for several monopole masses and fixed monopolium mass at $710$ GeV. Increasing the  binding energy  increases the cross section. Therefore the peak can be made as as small as desired by diminishing the binding energy. However, as will be shown in the next section, diminishing the mass of the monopole increases the $m-\bar{m}$ annihilation cross section and moreover makes the annihilation bump approach the monopolium peak. The same argument to discard Dirac coupling could be applied to disregard very small monopolium peaks.

We analyze  in the next section the onset on monopole-antimonopole annihilations and see that low expectations for these restrict further the monopole mass.

\section{Monopole-antimonopole annihilation}

It is natural to think that the enormous strength and long range of the  monopole interaction leads to the annihilation of the pair  into photons very close to the production point. Thus one should  look for monopoles through their annihilation into highly energetic photons, a channel for which LHC detectors have been optimized.
 
\begin{figure}[htb]
\begin{center}
\includegraphics[scale= 0.6]{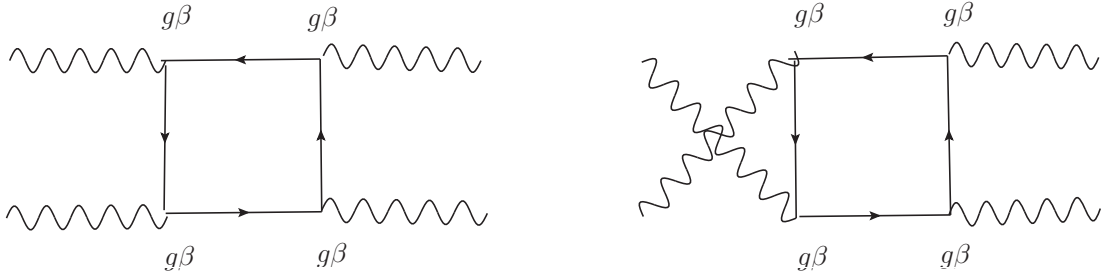}
\caption{ Elementary processes for  monopole-antimonopole production and annihilation into photons.}
\label{mmannihilation} 
\end{center}
\end{figure}

In order to calculate the annihilation into photons we  assume that our effective theory, a technically convenient modification of Ginzburg and Schiller's, agrees with QED at one loop order with a $\beta$ coupling.  Therefore we apply light-by-light scattering with the appropriate modifications as shown in Fig.~\ref{mmannihilation} and discussed in detail in ref. \cite{Epele:2012jn}. It became clear there that close to threshold the cross section is quite isotropic and  away from threshold the forward cross section, which is very difficult to measure, is  larger than the right-angle one.  Since the detectors cannot detect all of the photons coming out, we  take the right-angle cross section as a conservative indication of the magnitudes to be expected.

\begin{figure}
[ptb]
\begin{center}
\includegraphics[
height=2.5in,
width=4.in
]{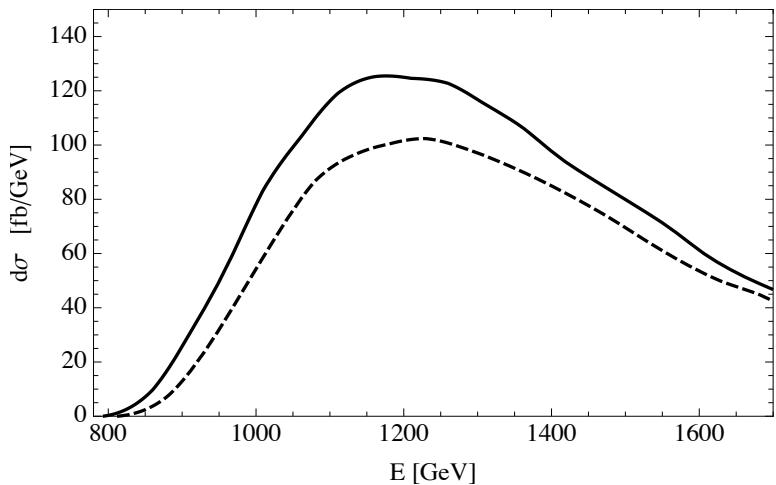}%
\caption{The 2 $\gamma$ cross section  for $m-\bar{m}$  annihilation in the $\beta$-coupling scheme as a function of the photon center of mass energy for a monopole mass of $380$ GeV (solid) and of $390$ GeV (dashed).}%
\label{mm380-390}%
\end{center}
\end{figure}

We  analyze the appearence of monopole-antimonopole after the monopolium enhancement following the steps of our previous work  \cite{Epele:2012jn}. In Fig. \ref{mm380-390} we show the annihilation cross section for two different masses. The cross section diminishes with increasing monopole mass. The maximum of the cross-section will occur far away from the threshold when $\beta$ approaches unity but it will be well below the $g$-coupling values since the energy denominator has already killed the maximum of the $g$-coupling strength when this happens. In any case, the results far away from threshold, must be taken with a grain of salt since our approximations might start to fail and moreover the angular dependence, not considered here, becomes important. 

\begin{figure}
[ptb]
\begin{center}
\includegraphics[
height=2.5in,
width=4in
]%
{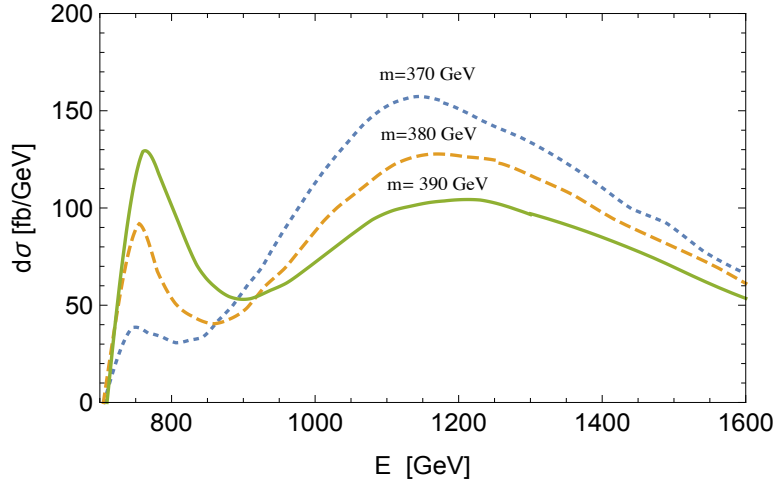}%
\caption{The 2 $\gamma$ cross section  for monopolium decay and monopole-antimonopole annihilation for the $\beta$-coupling scheme as a function of the two photon center of mass energy for three monopole masses.}%
\label{change}%
\end{center}
\end{figure}

It must be recalled that by keeping the monopolium mass fixed increasing the binding is equivalent to increasing the monopole mass. Thus monopolium annihilation and monopole-antimonopole annihilation cross sections vary in opposite way as shown ing Fig \ref{change}. This tightens the fixing of the monopole mass  since a mass which leads to a small monopolium annihilation cross section will lead to a large monopole-antimonopole cross section and viceversa. Once the mass of the monopole is fixed the annihilation cross section is fixed, there is no additional freedom in the model used. 

\begin{figure}
[ptb]
\begin{center}
\includegraphics[
height=2.5in,
width=4in
]%
{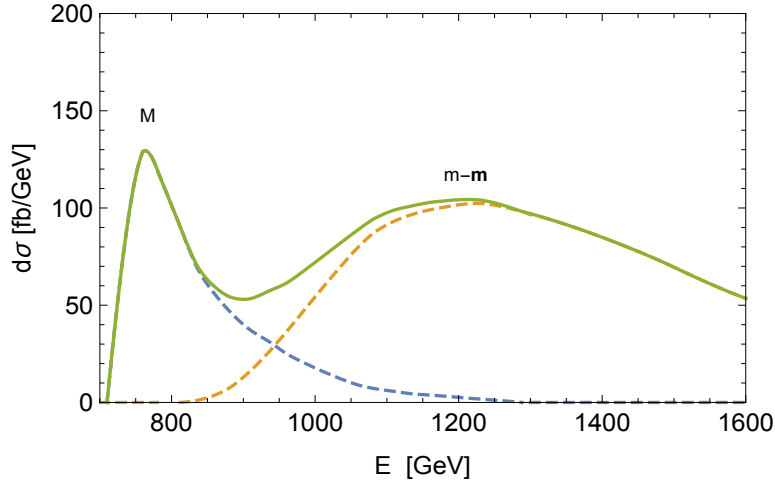}%
\caption{The 2 $\gamma$ cross section  for monopolium decay and monopole-antimonopole annihilation for the $\beta$-coupling scheme as a function of the two photon center of mass energy. We use a monopole mass of $390$ GeV, and a monopolium mass of $710$ GeV.}%
\label{fulldetail}%
\end{center}
\end{figure}

In Fig \ref{fulldetail}  we show the result of our calculation  for a $390$  GeV mass monopole  with a $710$ GeV mass monopolium in the $\beta$-coupling scheme . Due to the momentum dependent $\beta$-coupling the bump of the resonance is displaced by $40-50$ GeV from its mass. The 2 $\gamma$ annihilation signal above monopole-antimonopole threshold rises slowly and  is a very broad bump, whose maximum cross section is in this case about $100$ fm/GeV.  Smaller monopole masses and smaller binding energies lead to smaller cross sections for monopolium, while on the contrary for smaller masses the cross section for $m-\bar{m}$ annihilation is bigger,  as shown before. This different behavior is associated  to the binding energy of monopolium.

\section{Conclusions}

We have analyzed in this paper the possibility that the $750$ GeV enhancement could be a monopolium state. In order to have that possibility the mass of the monopole has to be low by conventional standards $< 400$ GeV. This low mass leads to the immediate conclusion that a monopole with Dirac coupling  cannot be  the origin of the enhancement. We have therefore limited our discussion to the so called $\beta$ coupling scheme.  In the $\beta$ scheme  the small width to $2 \gamma$ allows for a way out of the mass slavery. For a two photon branching ratio of a few percent, in agreement with most analysis \cite{Harland-Lang:2016qjy,Fichet:2015vvy,Csaki:2016raa}, we obtain a cross sections below $100$ fm/GeV signalling a monopolium state at $700-710$ GeV.

The most recent analysis \cite{Aaboud:2016tru,Khachatryan:2016hje} tend to lower the expectation on the 2 $ \gamma$ enhancement. If this were so, the lesson to be learned from our analysis is that the existence of a monopolium state implies automatically a threshold for monopole-antimonopole production relatively close by. Therefore  this monopolium peak should be accompanied by a $m-\bar{m}$ broad bump. The existence of the latter puts additional restrictions on the model parameters due to the magnitude of the annihilation cross section. In the present case in order to keep both cross sections low, the monopole mass must be in the range $370-400$ GeV.
 
Lower cross section to the ones shown in this paper for these low masses can only be achieved by increasing the total decay width above  $50$ GeV, and/or reducing the 2 $\gamma$ branching ratio even more. This mechanism would automatically reduce the monopolium cross section, would allow for larger  binding energies and larger monopole masses, and thus would  diminish  the $m-\bar{m}$ annihilation cross section furthermore.
 
A monopole scenario for the $750$ GeV enhancement  leads to a double bump, the lower can be a narrow peak associated with monopolium while the higher is a  broad bump associated to $m-\bar{m}$ annihilation into photons.  Such a structure will always arise for low mass monopoles ($< 500$ GeV). For higher mass monopoles  the bumps will broaden and the possibility of them merging together is high for reasonable binding energies, thus in that case one broad bump should be the most common feature \cite{Epele:2012jn}. 

Waiting expectantly for the data of the last LHC run we might conclude that  a $750$ GeV enhancement  followed by a soft rising background would be a plausible signal for monopole discovery.

After the completion of this paper we became aware of a non-perturbative lattice gauge theory calculation of monopolium and its eventual correlation with the $750$ GeV enhancement \cite{Barrie:2016wxf}.

\section*{Acknowledgements}
We would like to acknowledge very useful conversations with Luis Anchordoqui.
This work has been supported in part by MINECO (Spain) Grant. No. FPA2014-53631-C2-
1-P, GVA-PROMETEOII/2014/066 and SEV-2014-0398. 
This work was presented in the 5th Moedal Collaboration meeting  in June 2016.

\end{document}